\documentclass[aps,prb,twocolumn,showpacs,superscriptaddress]{revtex4}
\usepackage{amssymb}
\usepackage{amsfonts}
\usepackage{graphicx}
\begin{document}


\title{Evidence for line nodes in the energy gap of the overdoped Ba(Fe$_{1-x}$Co$_{x}$)$_{2}$As$_{2}$ from low-temperature specific heat measurements}

\author{Gang Mu}
\email[]{mugang@sspns.phys.tohoku.ac.jp} \affiliation{Department of
Physics, Graduate School of Science, Tohoku University, Sendai
980-8578, Japan}
\author{Jun Tang}
\affiliation{World Premier International Research Center, Tohoku
University, Sendai 980-8578, Japan}
\author{Yoichi Tanabe}
\affiliation{World Premier International Research Center, Tohoku
University, Sendai 980-8578, Japan}
\author{Jingtao Xu}
\affiliation{World Premier International Research Center, Tohoku
University, Sendai 980-8578, Japan}
\author{Satoshi Heguri}
\affiliation{Department of Physics, Graduate School of Science,
Tohoku University, Sendai 980-8578, Japan}
\author{Katsumi Tanigaki}
\email[]{tanigaki@sspns.phys.tohoku.ac.jp} \affiliation{Department
of Physics, Graduate School of Science, Tohoku University, Sendai
980-8578, Japan} \affiliation{World Premier International Research
Center, Tohoku University, Sendai 980-8578, Japan}

\begin{abstract}
Low-temperature specific heat (SH) is measured on
Ba(Fe$_{1-x}$Co$_{x}$)$_2$As$_2$ single crystals in a wide doping
region under different magnetic fields. For the overdoped sample, we
find the clear evidence for the presence of the $T^2$ term in the SH
data, which is absent both for the underdoped and optimal doped
samples, suggesting the presence of line nodes in the energy gap of
the overdoped samples. Moreover, the field induced electronic
specific heat coefficient $\Delta\gamma(H)$ increases more quickly
with the field for the overdoped sample than the underdoped and
optimal doped ones, giving another support to our arguments. Our
results suggest that the superconducting gap(s) in the present
system may have different structures strongly depending on the
doping regions.
\end{abstract}

 \maketitle

\section{Introduction}

Plenty of efforts have been made on the study of the iron-pnictide
superconductors, since the discovery of superconductivity with $T_c$
= 26 K in LaFeAsO$_{1-x}$F$_x$.\cite{Kamihara2008} An important
issue concerning this new family of high-$T_c$ superconductors is
about the gap structure, which should provide clues to the
understanding of the microscopic pairing mechanism.\cite{Mazin
Review} Up to date, consensuses have been reached on several
systems, e.g. LaFePO, KFe$_2$As$_2$, BaFe$_2$(As$_{1-x}$P$_x$)$_2$,
and so on, that nodes exist on the gap
structure.\cite{LaFePO1,LaFePO2,LaFePO3,KFe2As2-1,KFe2As2-2,KFe2As2-3,Ba122-P}
However, experimental results gave rather contradicting conclusions
on this issue in other systems of the iron-pnictide
superconductors.\cite{WYL,Sato,ZhGQ,MuCPL1,Chien,HDing,Hashimoto,MuPRB,penetration-depth,HeatTransport,ARSH}
One problem here may come from the different qualities of the
samples studied by different groups. Recently, annealing of the
single crystals in the 122 phase was reported to be an effective
route to improve the quality of the
samples.\cite{anneal1,anneal2,anneal3,anneal4} Especially, it was
found that annealing can considerably decrease the residual SH
coefficient $\gamma_0$ in the superconducting state and suppress the
Schottky anomaly in low temperature, which may suggest that fewer
impurities exist in the annealed samples.\cite{anneal4} This
supplies a better platform both for investigating the intrinsic
properties of the samples and for analyzing the SH data. Moreover,
some groups also point out that the gap structure shows a strong
evolution with the doping
concentration.\cite{penetration-depth,HeatTransport,anneal4}

Specific heat is one of the powerful tools to measure the
quasiparticle density of states (DOS) at the Fermi level, so as to
detect the information about the gap structure. The variation of
electronic SH versus temperature and magnetic field can be rather
different in the superconducting materials with different gap
structures. From textbook knowledge we know that the low-temperature
electronic SH $C_{el}$ for a superconductor with an isotropic gap
should have an exponential temperature dependence, namely $C_{el}
\propto e^{-\Delta_0/k_BT}$, where $\Delta_0$ is the magnitude of
the energy gap. However, for a superconductor with nodal gap(s), a
power law dependence of temperature for $C_{el}$ has been predicted
theoretically: $C_{el} \propto T^2$ for the gap with line nodes and
$C_{el} \propto T^3$ for point nodes.\cite{review1,review2} In fact,
the quadratic term has been observed in cuprate superconductors,
suggesting that line nodes exist in the gap
function.\cite{T2-1,T2-2,T2-3} In the mixed state, the magnetic
vortices in superconductors will induce depairing effect within and
outside the vortex cores, leading to the localized and delocalized
quasiparticle DOS, respectively. In general, the field-induced
electronic SH coefficient will show a rapid increase with the
increase of magnetic field in the low field region when the system
has the energy gap(s) with a rather small minimum
value.\cite{Vol1,Vol2,LSCO,MgB2}

For the iron-pnictide superconductors, it is rather difficult to get
the pure electron contribution from the total SH, since the upper
critical field is very high and the normal state can't be obtained
by suppressing the superconductivity using a magnetic field.
Nevertheless, many methods have been used by some groups to subtract
the phonon contribution from the SH data, which however, always
inevitably brings about uncertainties.\cite{MuPRB,Keimer,Ronning} In
this paper, we report the clear evidence of the presence of the
$T^2$ term in electronic SH of the overdoped
Ba(Fe$_{1-x}$Co$_{x}$)$_{2}$As$_{2}$ single crystals from the raw
data, which is consistent with the prediction of superconductors
with line nodes. Moreover, the field induced electronic SH
coefficient $\Delta\gamma(H)$ increases very quickly with magnetic
field for the overdoped sample, suggesting a rather small minimum
value of the energy gap. This further confirms the conclusion that
we stated above. These behaviors give a sharp comparison with that
observed in the underdoped and optimal doped samples, where a rather
small anisotropy is implied in the gap structure.

\begin{figure}
\includegraphics[width=9.5cm]{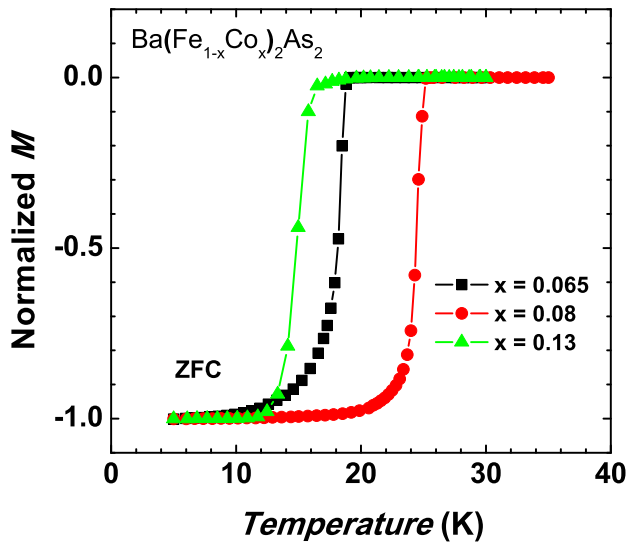}
\caption {(color online) Temperature dependence of dc magnetization
for three Ba(Fe$_{1-x}$Co$_{x}$)$_{2}$As$_{2}$ samples with $x$ =
0.065, 0.08, and 0.13, which are determined to be in the underdoped,
optimal doped, and overdoped regions, respectively. The data are
collected with field $H$ = 10 Oe using the zero field cooling (ZFC)
process. The curves are normalized by the magnetization data
obtained at 5 K. } \label{fig1}
\end{figure}

\section{Experimental Details and Sample Characterization}

The Ba(Fe$_{1-x}$Co$_{x}$)$_{2}$As$_{2}$ single crystals were grown
by the self-flux method.\cite{LFang} The as-grown samples were
annealed under high vacuum at 800 $^o$C for 20 days. The samples for
the SH measurement have typical dimensions of 2.5 $\times$ 1.5
$\times$ 0.2 mm$^{3}$. The dc magnetization measurements were done
with a superconducting quantum interference device (Quantum Design,
MPMS7). The specific heat were measured with a Helium-3 system based
on the Quantum Design instrument physical property measurement
system (PPMS). We employed the thermal relaxation technique to
perform the specific heat measurements. The thermometer has been
calibrated under different magnetic fields beforehand. The external
magnetic field is applied perpendicular to the $c$ axis of the
single crystals.

The superconducting transitions of the single crystals are checked
by the dc magnetization measurements. In Fig.~1, we show the
temperature dependence of the dc magnetization data for
Ba(Fe$_{1-x}$Co$_{x}$)$_{2}$As$_{2}$ samples with nominal doping
contents $x$ = 0.065, 0.08 and 0.13. The data were collected under a
dc field of 10 Oe, which have been normalized by the values obtained
at 5 K. The sample with $x$ = 0.08 is found to be optimally doped
with the highest onset transition temperature $T_c^{\mathrm{onset}}
\approx 25.2$ K, which is about 1 K higher than that of the as-grown
sample. Accordingly, the samples with $x$ = 0.065 and 0.13 are in
the underdoped and overdoped regions, respectively.

\begin{figure}
\includegraphics[width=8.5cm]{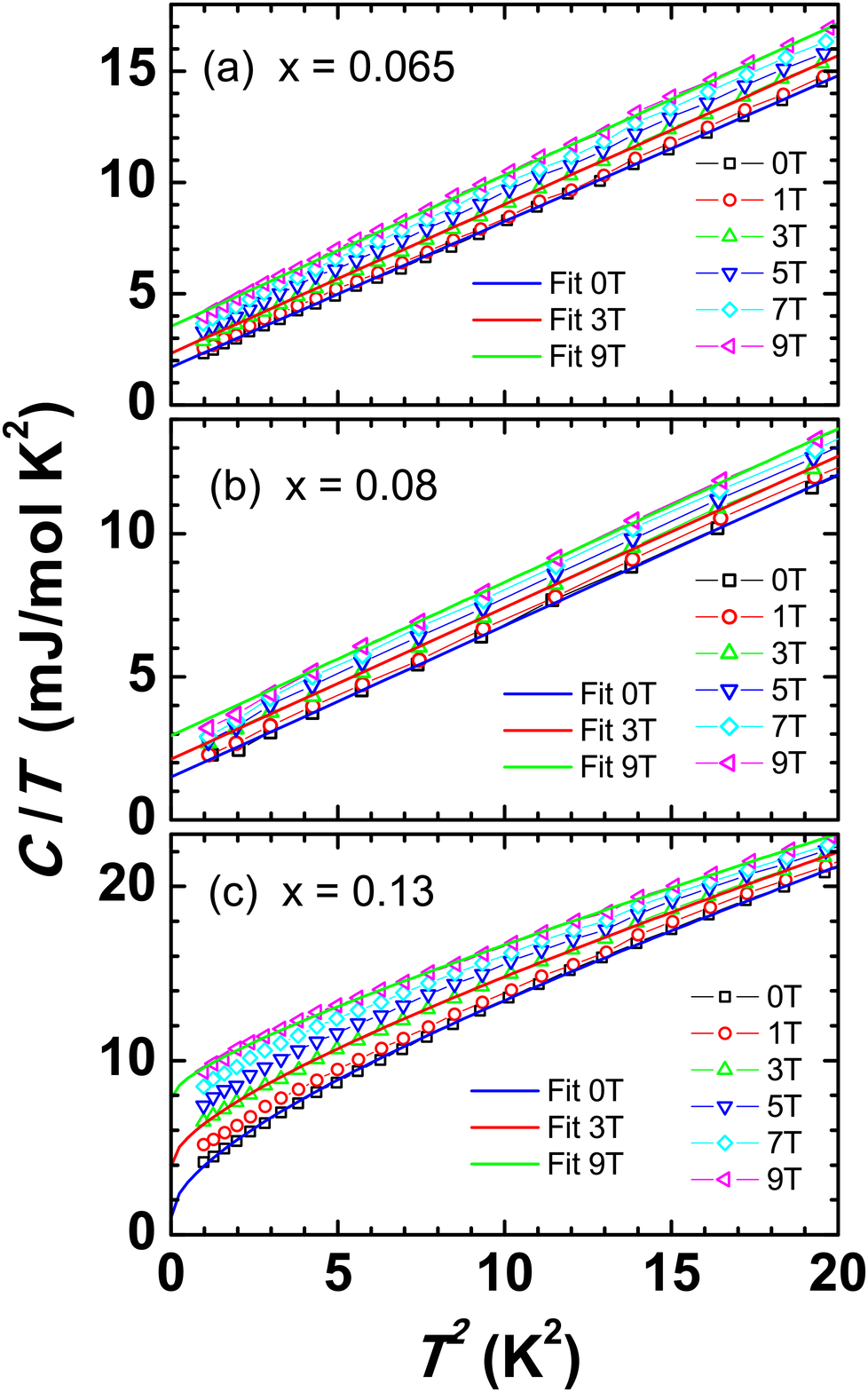}
\caption {(color online) The raw data of the SH for three
Ba(Fe$_{1-x}$Co$_{x}$)$_{2}$As$_{2}$ samples with $x$ = 0.065 (a),
$x$ = 0.08 (b), and $x$ = 0.13 (c) under different fields in the low
temperature region. The data are shown in $C/T$ vs $T^2$ plot. We
also show three solid lines of the theoretical fitting (see text) in
each figure. } \label{fig2}
\end{figure}

\section{Results and Discussion}

The raw data of SH for the three samples with $x$ = 0.065, 0.08, and
0.13 under different fields are plotted as $C/T$ vs $T^2$ in Fig.~2.
Here we focus on the behaviors of our data in the low temperature
region below 4.5 K. One can see clear different behaviors among the
three samples. For the underdoped and optimal doped samples, the
curves display clear linear tendency in this low temperature region,
which can actually be described by
\begin{equation}
C(T,H) = \gamma(H)T+\beta T^3 ,\label{eq:1}
\end{equation}
where $\gamma(H)$ is the electronic SH coefficient under a magnetic
field of $H$, and $\beta$ is the phonon SH coefficient. We show
three typical fitting curves based on equation (1) in Fig. 2(a) and
(b). From the fitting process, the residual SH coefficient under
zero field $\gamma_0$ ($\equiv \gamma(0)$) is determined to be 1.7
mJ/mol K$^2$ and 1.5 mJ/mol K$^2$ for the two samples, respectively.
A slight deviation from the linear behavior under high magnetic
fields in low temperatures may come from the influence of the
Schottky anomaly of the sample platform. We note that the linear
behavior is rather similar to the previous report.\cite{MuCPL2} For
the overdoped sample, however, the data show a clear negative
curvature rather than a straight line in all the curves up to 9 T,
in sharp contrast with the result of the underdoped and optimal
doped samples. We argue that this behavior suggests the presence of
the $T^2$ term in the electronic SH, which is expected for the
superconductors with line nodes in the energy
gap.\cite{review1,review2} Consequently, the data in Fig.~2(c) can
be described by
\begin{equation}
C(T,H) = \gamma(H)T+\alpha(H) T^2 + \beta T^3 ,\label{eq:2}
\end{equation}
with $\alpha(H)$ the coefficient of the $T^2$ term under the field
$H$. As shown by the solid lines, the fitting result is rather good.
We must note that the negative-curvature feature observed here can't
be attributed to the Schottky anomaly. In general, the behavior of
Schottky anomaly in SH should display a very strong evolution with
magnetic field and moreover a steep peak in the low-temperature
region under low fields should be observed, both of which are
missing in our experimental data. This gives a direct evidence for
the presence of line nodes in the energy gap of the overdoped
Ba(Fe$_{1-x}$Co$_{x}$)$_{2}$As$_{2}$. From the fitting, we find that
the residual term $\gamma_0 \approx$ 1.0 mJ/mol K$^2$ for this
sample, which is not larger than that of the optimal doped sample.
This is rather different from the results on the as-grown
samples.\cite{MuCPL2} The situation is reasonable since the
annealing process can dramatically suppress the number of the
impurity scatterers in the sample, as pointed by K. Gofryk et
al.\cite{anneal4} Accordingly, we know why the $T^2$ term can only
be detected in the annealed samples. As shown by the theoretical
work, the intraband impurity scattering tends to make the gap(s)
more isotropic, and may result in the removal of the low-energy
excitations.\cite{Mishra1,Mishra2}

\begin{figure}
\includegraphics[width=8.8cm]{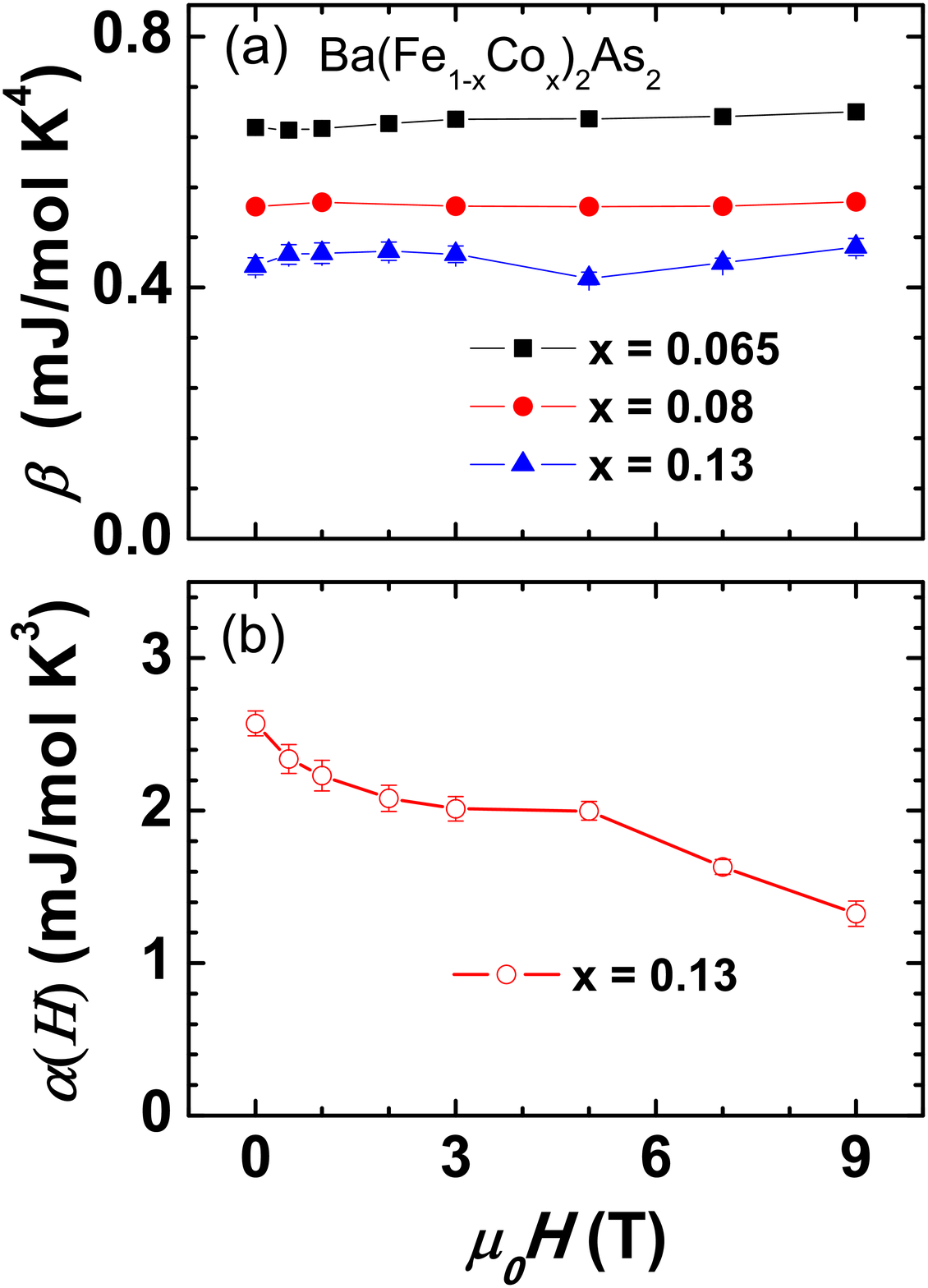}
\caption {(color online) Field dependence of the fitting parameters
$\beta$ and $\alpha(H)$ for the three samples. It is clear that
$\beta$ is almost independent of the field, while $\alpha(H)$ of the
overdoped sample decreases monotonously with the increase of the
field.} \label{fig3}
\end{figure}

The fitting parameters for the three samples, $\beta$ and
$\alpha(H)$, are shown in Fig.~3(a) and (b). One can see that the
value of $\beta$ is almost independent of field for the three
samples, which confirms the reliability of our analysis and fitting
process. Moreover, $\alpha(H)$ of the overdoped sample decreases
monotonously with the increase of magnetic field up to 9 T. For the
superconductors with a singlet d-wave gap (e.g. the cuprate
superconductors), it has been pointed out theoretically that the
$T^2$ term can only exist in moderate temperature regions
($\sqrt{H/H_{c2}}<<T/T_{c}<<1$) in the mixed state because of the
so-called Volovik effect.\cite{Vol1,Vol2,Vol3} Our observation here
may suggest that some segments of the line nodes are not affected by
the Volovik effect because the Fermi surface where they reside in is
perpendicular to the magnetic field. For the cuprate
superconductors, the value of $\alpha(H)$ under zero field was
estimated to be $\alpha(0)\sim \gamma_n/T_c$, with $\gamma_n$ the
electronic SH coefficient in the normal state.\cite{T2-2} Applying
this relation, $\alpha(0)$ is estimated to be about 1 mJ/mol K$^3$.
The value obtained from our data is about two times larger than this
estimation. This may suggest that the simple relation is not
suitable for the present multi-gap system, which needs more
investigations from the theoretical side. In addition, we note that
the values of $\beta$ decrease with the increase of the doping
content $x$. One possible origin of this behavior is that, for the
underdoped sample, a part of the $T^3$ term comes form the
contribution of the antiferromagnetic state, which coexists with the
superconducting state. There is another possibility. As mentioned
above, the electronic SH in a superconductor with point nodes can
also contribute the $T^3$ term. So the behavior of the data in
Fig.~3(a) may be interpreted in terms of the presence of point nodes
in the energy gap of the underdoped and optimal doped samples.
However, this argument is rather incompatible with the behavior
observed in Fig.~4, where the electronic SH coefficient increases
very slowly with the magnetic field both for the underdoped and
optimal doped samples (see below). One natural argument is that the
phonon SH is slightly different among the samples with different
doping contents.

\begin{figure}
\includegraphics[width=9.8cm]{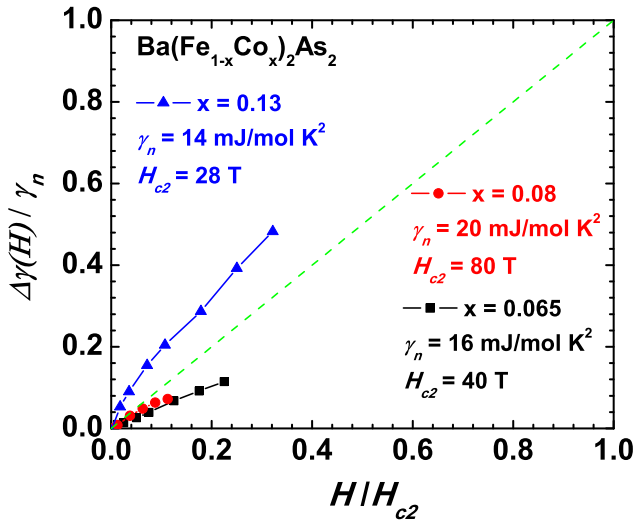}
\caption {(color online) The magnetic field dependence of the low
temperature electronic SH coefficient plotted as
$\Delta\gamma(H)/\gamma_n$ versus $H/H_{c2}$. The values of
$\gamma_n$ and $H_{c2}$ for the three samples are estimated based on
the result of other groups.\cite{anneal4,Hc2-1,Hc2-2} The green
dashed line is a guide for eyes, which shows the theoretical curve
for the samples with an isotropic superconducting gap. }
\label{fig4}
\end{figure}

The field dependence of the electronic SH coefficient
$\Delta\gamma(H) = \gamma(H)-\gamma(0)$ for the three samples is
shown in Fig.~4. The data are plotted as $\Delta\gamma(H)/\gamma_n$
versus $H/H_{c2}$, where $H_{c2}$ is the upper critical field. The
values of $\gamma_n$ and $H_{c2}$ are estimated according to the
results of other reports\cite{anneal4,Hc2-1,Hc2-2} and shown in this
figure. We must point out here that the uncertainty of the estimated
values of $\gamma_n$ and $H_{c2}$ will not affect the conclusions
deduced in the following. In this figure we also display a
theoretical curve for a conventional superconductor with an
isotropic superconducting gap by the green dashed line. One can see
that different behaviors of the three samples in different doping
regions are distinct. The data of the optimal doped sample seems
rather close to the green line under low magnetic fields, and even
evolves to be below it when the field increases. The underdoped
sample behaves similarly to the optimal doped sample. On the other
hand, the data of the overdoped sample increase more quickly and are
clearly above the green line. The behavior of the overdoped sample
is quite similar to those reported by others, although they have the
magnetic field along the $c$-axis of the crystals.\cite{anneal4} We
note here that it is not easy to describe the data in Fig.~4 using a
simple formula, because the present system was found to display
multi-band and even multi-gap features. Nevertheless, from the
present data we can draw the conclusion qualitatively that the
energy gap(s) of the overdoped sample has rather large anisotropy,
compared with that of the underdoped and optimal doped samples.
Obviously, this argument is consistent with our above-mentioned
conclusion that line nodes exist in the energy gap of the overdoped
sample.

Our results suggest that the gap structure of the
Ba(Fe$_{1-x}$Co$_{x}$)$_{2}$As$_{2}$ system shows a clear evolution
with the doping content $x$. Actually, doping dependence of the gap
structure has been detected by other experiments, including
penetration depth, heat transport, specific heat, point-contact
Andreev reflection, and so
on.\cite{penetration-depth,HeatTransport,anneal4,CRen} These results
share the same tendency with the present work that nodes should
exist in the overdoped samples, although there is discrepancy
concerning the underdoped and optimal doped regions. The detailed
mechanism of this behavior is still vague at this time. Even so, we
can unambiguously obtain important implications that the emergence
of line nodes in the overdoped sample is related to the evolution of
the topology of the Fermi surface and the pairing interaction with
the doping of electrons into the system, rather than the intrinsic
symmetry of the energy gap.

\section{Concluding remarks}

In summary, we studied the low-temperature specific heat on the
Ba(Fe$_{1-x}$Co$_{x}$)$_{2}$As$_{2}$ single crystals in a wide
doping region. Before measurements, the single crystals were
carefully annealed to improve the sample quality. We found a clear
evidence for the presence of $T^2$ term from the raw SH data on the
overdoped sample, which is absent both for the underdoped and
optimal doped samples, suggesting that line nodes must exist in the
energy gap of the overdoped sample. On the other hand, the
underdoped and optimal doped samples display rather small anisotropy
in the gap structure. Moreover, the field induced term
$\Delta\gamma(H)$ increased more quickly with magnetic field for the
overdoped sample than the underdoped and optimal ones, being
consistent with the above conclusions. Our results suggest that the
superconducting gap(s) in the present system may have different
structures depending on the different doping regions. Future
investigations on this issue on other systems of the iron-pnictide
superconductors are needed in the future.

\begin{acknowledgments}
We acknowledge the discussions with Dr. Yue Wang. The research was
partially supported by Scientific Research on Priority Areas of New
Materials Science Using Regulated Nano Spaces, the Ministry of
Education, Science, Sports and Culture, Grant-in-Aid for Science,
and Technology of Japan. The work was partly supported by Tohoku
GCOE Program and by the approval of the Japan Synchrotron Radiation
Research Institute (JASRI). Gang Mu expresses special thanks to
Grants-in-Aid for Scientific Research from the Japan Society for the
Promotion of Science (JSPS) (Grant No. P10026).
\end{acknowledgments}


\begin{thebibliography}{99}

\bibitem{Kamihara2008} Y. Kamihara, T. Watanabe, M. Hirano, and H. Hosono, J. Am. Chem. Soc. {\bf130}, 3296 (2008).
\bibitem{Mazin Review} P. J. Hirschfeld, M. M. Korshunov, I. I. Mazin, arXiv:1106.3712 (2011).
\bibitem{LaFePO1} J.D. Fletcher, A. Serafin, L. Malone, J. Analytis, J-H Chu, A.S. Erickson, I.R. Fisher, and A. Carrington, Phys. Rev. Lett. {\bf102}, 147001 (2009).
\bibitem{LaFePO2} M. Yamashita, N. Nakata, Y. Senshu, S. Tonegawa, K. Ikada, K. Hashimoto, H. Sugawara, T. Shibauchi, and Y. Matsuda, Phys. Rev. B {\bf80}, 220509(R) (2009).
\bibitem{LaFePO3} C. W. Hicks, T. M. Lippman, M. E. Huber, J. G. Analytis, J. H. Chu, A. S. Erickson, I. R. Fisher, and K. A. Moler, Phys. Rev. Lett. {\bf103}, 127003 (2009).
\bibitem{KFe2As2-1} J. K. Dong, S. Y. Zhou, T. Y. Guan, H. Zhang, Y. F. Dai, X. Qiu, X. F. Wang, Y. He, X. H. Chen, and S. Y. Li, Phys. Rev. Lett. {\bf104}, 087005 (2010).
\bibitem{KFe2As2-2} H. Fukazawa, Y. Yamada, K. Kondo, T. Saito, Y. Kohori, K. Kuga, Y. Matsumoto, S. Nakatsuji, H. Kito, P. M. Shirage, K. Kihou, N. Takeshita, C. H. Lee, A. Iyo, and
H. Eisaki, J. Phys. Soc. Jpn. {\bf78}, 083712 (2009).
\bibitem{KFe2As2-3} K. Hashimoto, A. Serafin, S. Tonegawa, R. Katsumata, R. Okazaki, T. Saito, H. Fukazawa, Y. Kohori, K. Kihou, C. H. Lee, A. Iyo, H. Eisaki, H. Ikeda,
Y. Matsuda, A. Carrington, and T. Shibauchi, Phys. Rev. B {\bf82}, 014526 (2010).
\bibitem{Ba122-P} M. Yamashita, Y. Senshu, T. Shibauchi, S. Kasahara, K. Hashimoto, D. Watanabe, H. Ikeda, T. Terashima, I. Vekhter, A. B. Vorontsov, Y. Matsuda
, arXiv: 1103.0885 (2011).
\bibitem{WYL} Y. Wang, L. Shan, L. Fang, P. Cheng, C. Ren, and H. H. Wen, Supercond. Sci. Technol. {\bf22}, 015018 (2009).
\bibitem{Sato} T. Sato, S. Souma, K. Nakayama, K. Terashima, K. Sugawara, T. Takahashi, Y. Kamihara, M. Hirano, and H. Hosono, J. Phys. Soc. Jpn.
{\bf77}, 063708 (2008).
\bibitem{ZhGQ} S. Kawasaki, K. Shimada, G. F. Chen, J. L. Luo, N. L. Wang, and Guo-qing Zheng, Phys. Rev. B {\bf78}, 220506(R) (2008).
\bibitem{MuCPL1} G. Mu, X. Zhu, L. Fang, L. Shan, C. Ren, and H. H. Wen, Chin. Phys. Lett. {\bf25}, 2221 (2008).
\bibitem{Chien} T. Y. Chen, Z. Tesanovic, R. H. Liu, X. H. Chen, and C. L. Chien, Nature (London) {\bf453}, 1224 (2008).
\bibitem{HDing} H. Ding, P. Richard, K. Nakayama, K. Sugawara, T. Arakane, Y. Sekiba, A. Takayama, S. Souma, T. Sato, T. Takahashi, Z. Wang, X. Dai, Z. Fang, G. F. Chen, J. L. Luo, and N. L. Wang, Europhys. Lett. {\bf83}, 47001 (2008).
\bibitem{Hashimoto}K. Hashimoto, T. Shibauchi, T. Kato, K. Ikada, R. Okazaki, H. Shishido, M. Ishikado, H. Kito, A. Iyo, H. Eisaki, S. Shamoto, and
Y. Matsuda, Phys. Rev. Lett. {\bf102}, 017002 (2009).
\bibitem{MuPRB} G. Mu, H. Q. Luo, Z. S. Wang, L. Shan, C. Ren, and H. H. Wen, Phys. Rev. B {\bf79}, 174501
(2009).
\bibitem{penetration-depth} C. Martin, H. Kim, R. T. Gordon, N. Ni, V. G. Kogan, S. L. Bud¡¯ko, P. C. Canfield, M. A. Tanatar, and R. Prozorov, Phys. Rev. B {\bf81}, 060505(R) (2010).
\bibitem{HeatTransport}  J. P. Reid, M. A. Tanatar, X. G. Luo, H. Shakeripour, N. Doiron-Leyraud, N. Ni, S. L. Bud'ko, P. C. Canfield, R. Prozorov, and Louis Taillefer, Phys. Rev. B {\bf82}, 064501 (2010).
\bibitem{ARSH} B. Zeng, G. Mu, H.Q. Luo, T. Xiang, I. I. Mazin, H. Yang, L. Shan, C. Ren, P.C. Dai, and H. H. Wen, Natature Communications, {\bf1}, 112 (2010).
\bibitem{anneal1} S. R. Saha, N. P. Butch, K. Kirshenbaum, and J. Paglione, Physica C {\bf470}, S379 (2010).
\bibitem{anneal2}  J. Gillett, S. D. Das, P. Syers, A. K. T. Ming, J. I. Espeso, C. M. Petrone, and S. E. Sebastian, arXiv: 1005.1330 (2010).
\bibitem{anneal3} C. R. Rotundu, B. Freelon, T. R. Forrest, S. D. Wilson, P. N. Valdivia, G. Pinuellas, A. Kim, J.-W. Kim, Z. Islam, E. Bourret-Courchesne, N. E. Phillips, and R. J. Birgeneau, Phys. Rev. B {\bf82}, 144525 (2010).
\bibitem{anneal4}  K. Gofryk, A. B. Vorontsov, I. Vekhter, A. S. Sefat, T. Imai, E. D. Bauer, J. D. Thompson, and F. Ronning, Phys. Rev. B {\bf83}, 064513 (2011).
\bibitem{review1} M. Sigrist and K. Ueda, Rev. Mod. Phys. {\bf63}, 239 (1991).
\bibitem{review2} N. E. Hussey, Advances in Physics, {\bf51}, 1685 (2002).
\bibitem{T2-1} N. Momono and M. Ido, Physica C {\bf264}, 311 (1996).
\bibitem{T2-2} S. J. Chen, C. F. Chang, H. L. Tsay, H. D. Yang, and J.-Y. Lin, Phys. Rev. B {\bf58}, 14753 (1998).
\bibitem{T2-3} R. A. Fisher, N. E. Phillips, A. Schilling, B. Buffeteau, R. Calemczuk, T. E. Hargreaves, C. Marcenat, K. W. Dennis, R. W. McCallum, and A. S. O'Connor, Phys. Rev. B {\bf61}, 1473 (2000).
\bibitem{Vol1} G. E. Volovik, JETP Lett. {\bf58}, 469 (1993).
\bibitem{Vol2} G. E. Volovik, JETP Lett. {\bf65}, 491 (1997).
\bibitem{LSCO} H. H. Wen, Z. Y. Liu, F. Zhou, J. W. Xiong, W. X. Ti, T. Xiang, S. Komiya, X. F. Sun, and Y. Ando, Phys. Rev. B {\bf70}, 214505 (2004).
\bibitem{MgB2} F. Bouquet, R. A. Fisher, N. E. Phillips, D. G. Hinks, and J. D. Jorgensen, Phys. Rev. Lett. {\bf87}, 047001 (2001).
\bibitem{Keimer} P. Popovich, A. V. Boris, O. V. Dolgov, A. A. Golubov, D. L. Sun, C. T. Lin, R. K. Kremer, and B. Keimer, Phys. Rev. Lett. {\bf105}, 027003 (2010).
\bibitem{Ronning} K. Gofryk, A. S. Sefat, M. A. McGuire, B. C. Sales, D. Mandrus, J. D. Thompson, E. D. Bauer, and F. Ronning, Phys. Rev. B {\bf81}, 184518 (2010).
\bibitem{LFang}  L. Fang, H. Luo, P. Cheng, Z. Wang, Y. Jia, G. Mu, B. Shen, I. I. Mazin, L. Shan, C. Ren, and H. H. Wen, Phys. Rev. B {\bf80} 140508(R) (2009).
\bibitem{MuCPL2} G. Mu, B. Zeng, P. Cheng, Z. S. Wang, L. Fang, B. Shen, L. Shan, C. Ren, and H. H. Wen, Chin. Phys. Lett. {\bf27}, 037402 (2010).
\bibitem{Mishra1} V. Mishra, G. Boyd, S. Graser, T. Maier, P. J. Hirschfeld, and D. J. Scalapino, Phys. Rev. B {\bf79}, 094512 (2009).
\bibitem{Mishra2} V. Mishra, A. Vorontsov, P. J. Hirschfeld, and I. Vekhter, Phys. Rev. B {\bf80}, 224525 (2009).
\bibitem{Vol3} N. B. Kopnin and G. E. Volovik, JETP Lett. {\bf64}, 690 (1996).
\bibitem{Hc2-1} M. Kano, Y. Kohama, D. Graf, F. Balakirev, A. S. Sefat, M. A. Mcguire, B. C. Sales, D. Mandrus, and S. W. Tozer, J. Phys. Soc. Jpn. {\bf78}, 084719 (2009).
\bibitem{Hc2-2} A. Yamamoto, J. Jaroszynski, C. Tarantini, L. Balicas, J. Jiang, A. Gurevich, D. C. Larbalestier, R. Jin, A. S. Sefat, M. A. McGuire, B. C. Sales,
 D. K. Christen, and D. Mandrus, Appl. Phys. Lett. {\bf94}, 062511 (2009).
\bibitem{CRen} C. Ren, Z. S. Wang, Z. Y. Wang, H. Q. Luo, X. Y. Lu, B. Sheng, C. H. Li, L. Shan, H. Yang, and H. H. Wen, arXiv:1106.2891 (2011).



\end{thebibliography}
\end{document}